\documentclass[fleqn,usenatbib]{mnras}

\usepackage{newtxtext,newtxmath}

\usepackage[T1]{fontenc}

\DeclareRobustCommand{\VAN}[3]{#2}
\let\VANthebibliography\thebibliography
\def\thebibliography{\DeclareRobustCommand{\VAN}[3]{##3}\VANthebibliography}

\usepackage{graphicx}	
\usepackage{amsmath}	

\newcommand\oiii    	{$\mathrm{\left[ O \textsc{iii}\right] }$}

\newcommand\mmsun	    {\rm{M}_{\odot}}
\newcommand\ha          {H$\upalpha$}

\newcommand\hb          {H$\upbeta$}
\newcommand\mstar       {$M_{\ast}$}
\newcommand\mmstar      {M_{\ast}}
\newcommand\mbh         {$M_{\mathrm{BH}}$}
\newcommand\mmbh        {M_{\mathrm{BH}}}
\newcommand\mbulge      {$M_{\mathrm{Bulge}}$}
\newcommand\mmbulge     {M_{\mathrm{Bulge}}}
\newcommand\lbol        {$L_{\mathrm{bol}}$}
\newcommand\mlbol       {L_{\mathrm{bol}}}
\newcommand\ledd        {$L_{\mathrm{Edd}}$}
\newcommand\mledd       {L_{\mathrm{Edd}}}
\newcommand\fedd        {$\lambda_{\mathrm{Edd}}$}
\newcommand\mfedd       {\lambda_{\mathrm{Edd}}}
\newcommand\reff        {$R_{e}$}

\newcommand\btotL       {$\lambda_\mathrm{B/tot}$}
\newcommand\mbtotL      {\lambda_\mathrm{B/tot}}
\newcommand\pbtotL      {$\lambda_\mathrm{pB/tot}$}
\newcommand\mpbtotL     {\lambda_\mathrm{pB/tot}}
\newcommand\htotL       {$\lambda_\mathrm{host/tot}$}

\newcommand\agntotL     {$\lambda_\mathrm{AGN/tot}$}
\newcommand\magntotL    {\lambda_\mathrm{AGN/tot}}

\newcommand\mrhoS       {\rho_{\mathrm{S}}}

\newcommand\mrhoP       {\rho_{\mathrm{P}}}

\newcommand\kpc         {\,\textrm{kpc}}
\newcommand\Mpc         {\,\textrm{Mpc}}
\newcommand\s           {\,\textrm{s}}
\newcommand\km          {\,\textrm{km}}
\newcommand\yr          {\,\textrm{yr}}
\newcommand\erg         {\,\mathrm{erg}}
\newcommand\micrometer  {\,\mu\textrm{m}}

\newcommand\HST         {\emph{HST}}
\newcommand\Euclid      {\emph{Euclid}}
\newcommand\galfit      {\texttt{GALFIT}}

\newcommand\eg          {\emph{e.g.},}
\newcommand\ie          {\emph{i.e.},}

\defcitealias{simmons2017}{S17}
\defcitealias{haring2004}{HR04}

\title[Merger-free AGN-hosts with GALFIT]
{Structural Decomposition of Merger-Free Galaxies Hosting Luminous AGNs}

\author[M. J. Fahey et al.]
{Matthew J. Fahey,$^{1}$\footnote[3]{First authorship is shared between Fahey and Garland}
Izzy. L. Garland,$^{2,1}$\footnote[1]{E-mail: garland@mail.muni.cz}\footnotemark[3]
Brooke. D. Simmons,$^{1}$
William C. Keel,$^{3}$
Jesse Shanahan,$^{4}$ \newauthor
Alison Coil,$^{4}$ 
Eilat Glikman,$^{5}$
Chris J. Lintott,$^{6}$
Karen L. Masters,$^{7}$
Ed Moran,$^{8}$
Rebecca J. Smethurst,$^{6}$  \newauthor
Tobias G\'{e}ron,$^{6,9}$
Matthew R. Thorne$^{1}$\\
$^{1}$Physics, Lancaster University, Lancaster LA1 4YB, UK\\
$^{2}$Department of Theoretical Physics and Astrophysics, Faculty of Science, Masaryk University, Kotl\'{a}\v{r}sk\'{a} 2, Brno, 611 37, Czech Republic\\
$^{3}$Department of Physics and Astronomy, University of Alabama, 206 Gallalee Hall, 514 University Blvd. Tuscaloosa, AL 35487-0324, USA\\
$^{4}$Center for Astrophysics and Space Sciences, University of California, San Diego, 9500 Gilman Dr, MC 0424, La Jolla, CA 92093-0424, USA \\
$^{5}$Department of Physics, Middlebury College, Middlebury, VT 05753, USA \\
$^{6}$Oxford Astrophysics, Department of Physics, University of Oxford, Denys Wilkinson Building, Keble Road, Oxford, OX1 3RH, UK\\
$^{7}$Departments of Physics and Astronomy, Haverford College, Lancaster Avenue, Ardmore, PA 19041 USA\\
$^{8}$Astronomy Department, Wesleyan University, Middletown, CT 06459, USA\\
$^{11}$Dunlap Institute for Astronomy and Astrophysics, University of Toronto, 50 St. George Street, Toronto, ON M5S 3H4, Canada}

\date{Accepted XXX. Received YYY; in original form ZZZ}

\pubyear{2024}

\begin{document}
\label{firstpage}
\pagerange{\pageref{firstpage}--\pageref{lastpage}}
\maketitle

\begin{abstract}
Active galactic nucleus (AGN) growth in disk-dominated, merger-free galaxies is poorly understood, largely due to the difficulty in disentangling the AGN emission from that of the host galaxy.
By carefully separating this emission, we examine the differences between AGNs in galaxies hosting a (possibly) merger-grown, classical bulge, and AGNs in secularly grown, truly bulgeless disk galaxies.
We use \galfit\ to obtain robust, accurate morphologies of 100 disk-dominated galaxies imaged with the \emph{Hubble Space Telescope}.
Adopting an inclusive definition of classical bulges, we detect a classical bulge component in $53.3 \pm 0.5$ per cent of the galaxies.
These bulges were not visible in Sloan Digital Sky Survey photometry, however these galaxies are still unambiguously disk-dominated, with an average bulge-to-total luminosity ratio of $0.1 \pm 0.1$.
We find some correlation between bulge mass and black hole mass for disk-dominated galaxies, though this correlation is significantly weaker in comparison to the relation for bulge-dominated or elliptical galaxies.
Furthermore, a significant fraction ($\gtrsim 90$ per cent) of our black holes are overly massive when compared to the relationship for elliptical galaxies.
We find a weak correlation between total stellar mass and black hole mass for the disk-dominated galaxies, hinting that the stochasticity of black hole-galaxy co-evolution may be higher disk-dominated than bulge-dominated systems.
\end{abstract}




\section{Introduction}

Strong correlations observed between supermassive black holes (SMBHs) and galaxy properties such as velocity dispersion \citep{magorrian1998, ferrarese2000, kormendy2001}, bulge stellar mass \citep{marconi2003, haring2004}, and total stellar mass \citep{Ferrarese2005,cisternas2011a, marleau2013}, have led to the general consensus that SMBH co-evolve with their host galaxies \citep[for a thorough review see][]{heckman2014}. Any theory of galaxy growth and evolution must endeavour to explain these correlations.

Galaxy mergers are a key process through which stars are transferred from rotation-supported galactic disks to dispersion-supported bulges \citep{walker1996, hopkins2012, welker2017}. Simulations of merging galaxies suggest that their gas content is subject to a significant gravitational torque \citep{hernquist1989, barnes1996}, and that this violent process involves substantial angular momentum transfer which funnels gas from kiloparsec-scale to parsec-scale orbits. This often allows increased volumes of gas to be accreted onto a central SMBH, triggering an active galactic nucleus \citep[AGN;][]{shlosman1989}. Thus, major mergers have been regarded as the primary mechanism through which a galaxy can grow both its stellar and black hole mass \citep{sanders1988, croton2006, hopkins2006, peng2007}.

If major mergers are indeed the primary driver of co-evolution, then SMBH properties should correlate almost exclusively with galaxy properties tied to merger histories. Some past studies support these findings \citep{kormendy2011}, and it has been found that luminous quasars seem preferentially hosted in ongoing mergers \citep{volonteri2006, urrutia2008, glikman2015, trakhtenbrot2017}. Studies also suggest AGN triggering due to major merger events \citep{treister2012, bessiere2014, gao2020}, with additional evidence suggesting AGN activity peaks post-merger \citep{schawinski2010, ellison2013}.

However, recent observational studies have cast increasing doubt on the relevance of major mergers for triggering AGN activity and facilitating SMBH-galaxy co-evolution. Vast numbers of X-ray detected and optically observed AGNs across cosmic time show no significant connection with merger-driven triggering and fuelling \citep{gabor2009, georgakakis2009, cisternas2011b}. Furthermore, moderate luminosity AGNs representing more typical rates of SMBH growth \citep{hasinger2005} show an almost identical trend: that there is no observational connection between the triggering and fuelling of AGNs by major mergers \citep{grogin2005, allevato2011, schawinski2011, kocevski2012, cisternas2015, rosario2015, goulding2017, marian2019, silva2021}.

In addition, cosmological simulations have suggested that whilst mergers may increase the rate of luminous AGNs, they do not play a significant role in driving SMBH growth \citep{McAlpine2020, smethurst2024}, with as little as 35 per cent of black hole growth since $z \sim 3$ traceable back to a major or minor merger \citep{martin2018}.

The morphology of a given galaxy can reveal its merger history -- disk-dominated, bulgeless galaxies have had no major mergers in their history since $z\sim2$ \citep{martig2012}. While it is possible for gas-rich disks to re-form following a major merger event \citep[see][]{hopkins2009, emsellem2011, pontzen2017, sparre2017}, a central classical bulge component remains as a relic in the post-merger structure. The minimum galaxy mass ratio required for a merger event to create a bulge is routinely cited as 1:10 \citep{walker1996, hopkins2012, tonini2016}, though theoretical studies have introduced some uncertainty into this number \citep{hopkins2009, brook2012, kannan2015}. 

Despite their visual similarity to classical bulges, pseudobulges likely grow through purely secular processes \citep{kormendy2004, kormendy2010}. Additional work has suggested that classical bulges may also form and grow through processes other than mergers \citep{scannapieco2009, sales2012, zolotov2015, bell2017, park2019, wang2019, guo2020}. This suggests that galaxies which are unambiguously disk-dominated in their morphologies have had an extremely calm baryon accretion history, evolving in the absence of major mergers since $z \sim 2$ \citep{martig2012}. 

Using conservative upper limits on bulge mass in a sample of disk-dominated galaxies, \citet*{simmons2017} show that AGN can not only exist in the absence of merger-grown morphological features, but SMBHs in disk-dominated galaxies can grow to larger masses than predicted by the \citet{haring2004} relationship between bulge stellar mass and SMBH mass. However, this study was done with SDSS imagery, which lacks the high-resolution that can be obtained with space-based imaging, motivating the \emph{Hubble Space Telescope} (\HST) observations discussed in Section \ref{sec:HSTSnapshotSurvey}.

\citet{smethurst2020} investigate possible powering mechanisms for their population of merger-free, disk-dominated AGNs. Having calculated a mean mass inflow rate of $1.01 \pm 0.14 \,\mmsun\,yr^{-1}$ they show that morphological features such as bars \citep{sakamoto1996, maciejewski2002, regan2004, lin2013} and spiral arms \citep{maciejewski2004, davies2009, schnorrmuller2014}, as well as smooth accretion of cold gas \citep{keres2005, sancisi2008} can match this required rate, and hence should be capable of sustainably fuelling an AGN. Furthermore, as morphological features such as bars and spiral arms are thought to be long lasting \citep{miller1979, sparke1987, donner1994, donghia2013, hunt2018}, these inflow rates can be sustained in the long term, implying purely secular processes can rapidly grow SMBHs.

Due to the relative rarity of merger-free galaxies hosting luminous AGNs, previously studied samples have routinely suffered from issues such as: being too small to statistically constrain black-hole galaxy relations dependent on merger-free pathways, making use of selection techniques that preclude sampling the full parameter space of galaxy and black-hole mass, and including obscured AGNs for which the only available SMBH mass estimation frameworks are highly uncertain. While recent work has vastly improved upon these issues, a lack of high resolution imaging has placed a fundamental upper limit on the extent to which any observed correlations can be constrained. Previous work has pointed to the importance of parametric image fitting for decomposing galaxy morphology \citep{kim2008, simmons2008, pierce2010}, and so high quality imagery, such as that from \HST, is required for an independent assessment of host morphology and to quantify bulge contribution.

We present a sample of 100 luminous, unobscured AGNs hosted in disk-dominated galaxies imaged with \HST. We make use of the 2D paramteric image fitting software \galfit\ \citep{peng2002, peng2010} to classify each morphology component individually and make clear distinctions between classical bulges and pseudobulges. We investigate the extent to which galaxies hosting bulges differ from the bulgeless population in terms of their AGN properties, and explore the mechanisms by which these differences might be explained.

Section \ref{sec:ObservationalData} describes the selection of disk-dominated galaxies hosting luminous AGNs and the observational \HST\ data. Section \ref{sec:StructuralDecomposition} describes the structural decomposition of these sources using \galfit, and the schemes used for component classification. Section \ref{sec:AdditionalParameters} outlines the calculation of luminosity ratios, galaxy stellar masses, and black hole masses to better inform our analysis. Section \ref{sec:BarFraction} investigates the barred fraction of our galaxies in detail. Section \ref{sec:GalBHMassRels} investigates black hole--galaxy relations and black hole--galaxy co-evolution. Our findings are summarised in Section \ref{sec:Summary}.

Throughout this paper, all cross-matched catalogues use the nearest positional match within 3\arcsec. We use the AB magnitude system \citep{oke1983}, and adopt a flat $\Lambda$CDM cosmology using \textsc{astropy} \citep{robitaille2013, pricewhelan2022}, with $H_0 = 70.0 \km\,\s^{-1}\Mpc^{-1}$, $\Omega_{\mathrm{M}} = 0.3$ and $\Omega_{\Lambda}$ = 0.7.

\section{Observational Data}\label{sec:ObservationalData}
We utilise the sample first compiled in \citet[hereafter \citetalias{simmons2017}]{simmons2017}  who investigated BH growth in galaxies that evolved in the absence of major mergers using SDSS imagery.
The authors selected an initial unobscured AGN sample using W2R \citep{edelson2012}, which consists of infrared data from the \emph{Wide-field Infrared Survey Explorer} \citep[\emph{WISE}:][]{wright2010} and the Two-Micron All Sky Survey \citep[2MASS][]{skrutskie2006}, and X-ray data from the \emph{ROSAT} All Sky Survey \citep[RASS:][]{voges1999}. From this sample of 4,316 unobscured AGNs, 1,844 have SDSS sources within 3\arcsec. A single expert classifier (BDS) used SDSS photometry to select the galaxies with no evidence of a bulge component, but with features commonly found in disks, such as bars and spiral arms. This resulted in 137 disk-dominated galaxies hosting unobscured, Type 1 AGNs.

We refer to \citetalias{simmons2017} for further information on the selection and analysis of this initial 137 source sample.

\subsection{\HST\ Snapshot Survey} \label{sec:HSTSnapshotSurvey}

\begin{figure*}
    \centering
    \includegraphics[width=\textwidth]{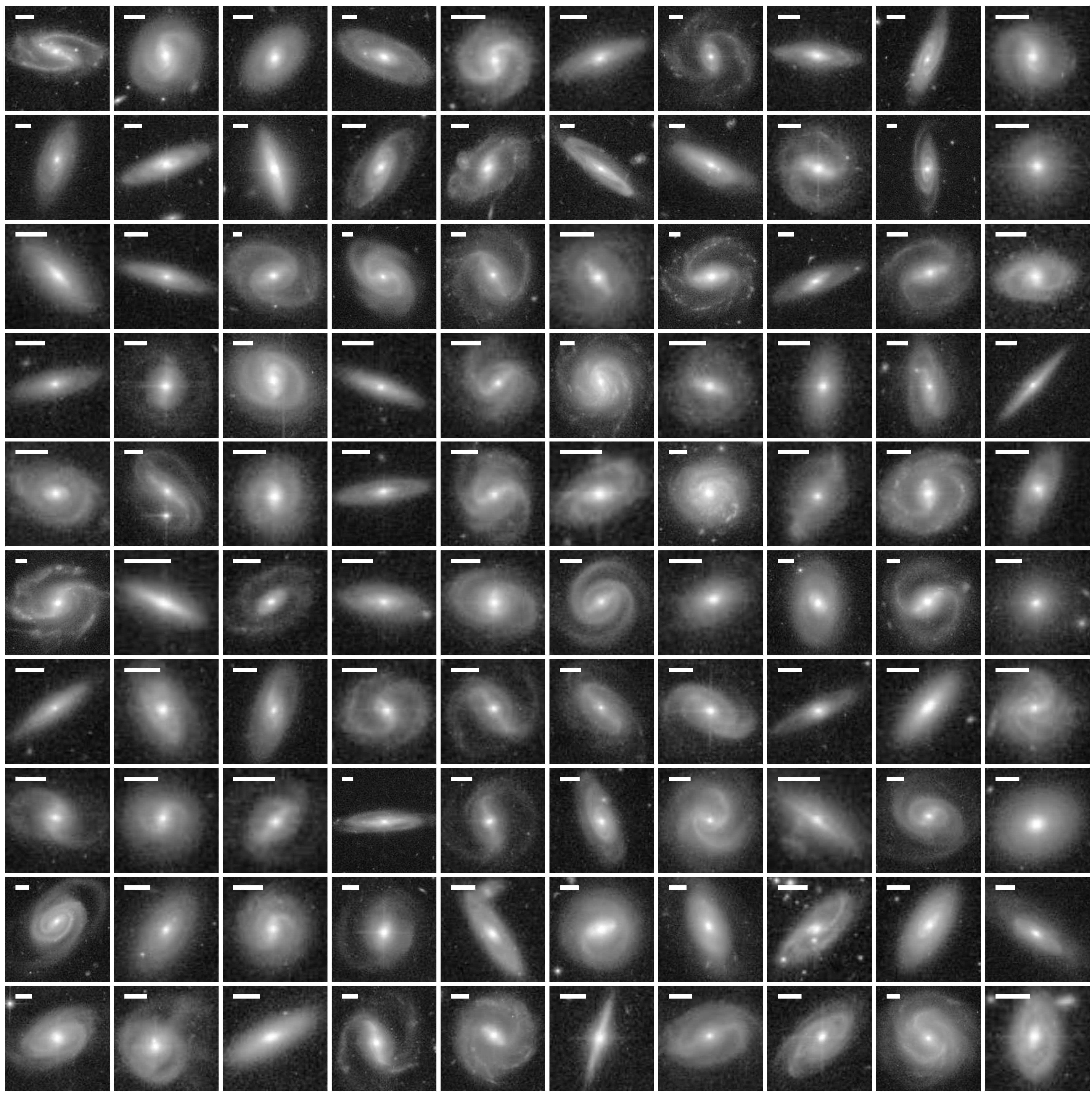}
    \caption{Postage stamp \HST\ images of the sample of 100 disk-dominated galaxies with unobscured, luminous AGNs. Sample selection is described in Sections \ref{sec:ObservationalData} and \ref{sec:HSTSnapshotSurvey}. Images are displayed in the order they were imaged by \HST. Scale bars in each panel show 5\arcsec. These 100 sources comprise our sample for analysis.}
    \label{fig:HSTimages}
\end{figure*}

Previous morphological decomposition of the \citetalias{simmons2017} sample was limited in accuracy by the comparatively high point-spread function (PSF) of SDSS. This may have led to the AGN luminosity contributing to the measured bulge luminosity. Furthermore, apart from two objects with dedicated Isaac Newton Telescope (INT) imagery, all of the bulge-to-total mass ratio (\btotL) and bulge stellar mass (\mbulge) estimates of \citetalias{simmons2017} are considered conservative upper limits due to the presence of the luminous AGN. These limitations motivated a follow-up survey performed using \HST\ in order to obtain much more detailed imagery.

The \HST\ snapshot survey (programme ID HST-GO-14606 PI: Simmons) was performed with broad band imaging in the F814W, F850LP, and F775W filters using the Advanced Camera for Surveys (ACS). \HST's ACS module is chosen due its very stable and extremely well-modelled PSF, which is ideal for the accurate separation of host galaxy from AGN. Due to the limitations of a snapshot programme, sources with less clear morphology in SDSS, as determined visually, were prioritised. Filters were chosen so as to minimise the contribution from the AGN by avoiding emission lines such as \oiii, \ha\ and \hb. We also selected filters to image the reddest possible broadband rest-frame optical filter that is not significantly affected by dust in order to capture the typically redder bulge emission required for accurate host galaxy decomposition. F775W is chosen for observations of targets at $z < 0.06$, F814W for targets at $0.06 < z < 0.08$, and F850LP for targets at $z > 0.08$. 101 galaxies were imaged in total: 3 in F775W, 38 in F814W, and 60 in F850LP.

Each source was imaged four times: two short exposures to capture an unsaturated nuclear PSF, and two long exposures to obtain the extended emission to an acceptable depth. This resulted in a typical exposure time of 40 minutes across the four exposures. ACS/WFC subarrays were chosen to ensure substantial sky background was imaged whilst minimising readout time. The data were then reduced using the standard pipeline\footnote[1]{Some manual steps were required during the data reduction to account for the subarrays, but these steps have since been incorporated into the standard pipeline.}, accounting for CCD charge diffusion correction, cosmic ray removal using \texttt{lacosmic} \citep{vandokkum2001} and combination of the long exposures into a single science exposure.

The reduced images have a pixel scale of 0.05\arcsec\ per pixel, corresponding to a physical pixel scale of $0.041\kpc$, $0.12\kpc$, and $0.19\kpc$ at the minimum ($z = 0.042$), median ($z = 0.129$), and maximum ($z = 0.242$) redshift of the sample respectively.

One galaxy is excluded due to failed guide star acquisition. As such, the 100 remaining galaxies imaged by \HST\ comprise our sample for analysis throughout the rest of this paper. Postage stamp images of these 100 galaxies are shown in Fig. \ref{fig:HSTimages}.

\section{Structural Decomposition} \label{sec:StructuralDecomposition}
When inspecting galaxy images, previous work highlights the difficulty in distinguishing between point sources due to an AGN, and the presence of a classical galaxy bulge \citep{simmons2008}.  Hence, for a quantitative and robust decomposition of galaxy morphology, parametric fitting is necessary. This allows for both an independent assessment of host morphology, and quantitative separation between disk and bulge, providing constraints on the contribution of the bulge to the overall flux.

\subsection{Parametric fitting with \galfit}
We use the two-dimensional parametric image fitting program \galfit\ \citep{peng2002, peng2010} to simultaneously model the unresolved nucleus and extended galaxy emission for each of the galaxies in our sample using the \HST\ images obtained in Section \ref{sec:HSTSnapshotSurvey} \citep{dunlop2003, gabor2009, simmons2011, schawinski2011, schawinski2012}.

We initially employ a batch fitting procedure to fit each source three times. Firstly, the source is modelled with a combination of a single S\'{e}rsic profile \citep[with varying S\'{e}rsic index;][]{sersic1968} and a central point source. Secondly, the source is modelled with a single S\'{e}rsic profile, an exponential disk profile, and a central point source. Thirdly, the source is modelled with a single S\'{e}rsic profile, a classical \citet{devaucouleurs1953} bulge profile, and a central point source.

Initial parameters (magnitude, radius, axis ratio and position angle) were either drawn from the SDSS catalogue where available, or estimated based upon inspection of the \HST\ images.
We set the initial host S\'{e}rsic index to $n = 2.5$ (so as to avoid favouring either an exponential disk, $n = 1$, or a classical bulge, $n = 4$) and allow it to vary. The exponential disk and bulge profile in the second and third fits are initialised in a similar way, as S\'{e}rsic components with indices fixed to $n = 1$ and $n = 4$ respectively. The purpose of these initial fits is to converge on the centroid positions of each of the galaxy components.
During the batch fitting procedure we also fix the sky background to an independently determined value for each nearby source.
Any nearby object whose light profile is not impinging on the central galaxy is masked from the fit. In the case where nearby bright stars or companions galaxies impinge on the central galaxy, the relevant object is fit with its own component and noted as a `companion' (discussed further in Section \ref{sec:ComponentClassification}). The aim of the fit is to neither over nor under subtract the galaxy’s central region.

The residuals of these three initial fits are inspected for each galaxy, and the best fit (based on visual inspection) is refined via a further manual fitting process.
During the manual fitting process, additional model components were included based upon iterative visual inspection of the fit residuals, with the same goal of neither over nor under subtracting the central region. Great care was taken to ensure the chosen galaxy best fit contains component parameters which are physically reasonable.
These additional model components can take any form allowed by \galfit, but were typically additional S\'{e}rsic components.
In some cases, these S\'{e}rsic components require modification to best fit galaxy morphologies, such as truncation to model a strong galactic bar, or power-law rotation to model bright spiral-arms.

\begin{figure}
    \centering
    \includegraphics[width=\columnwidth]{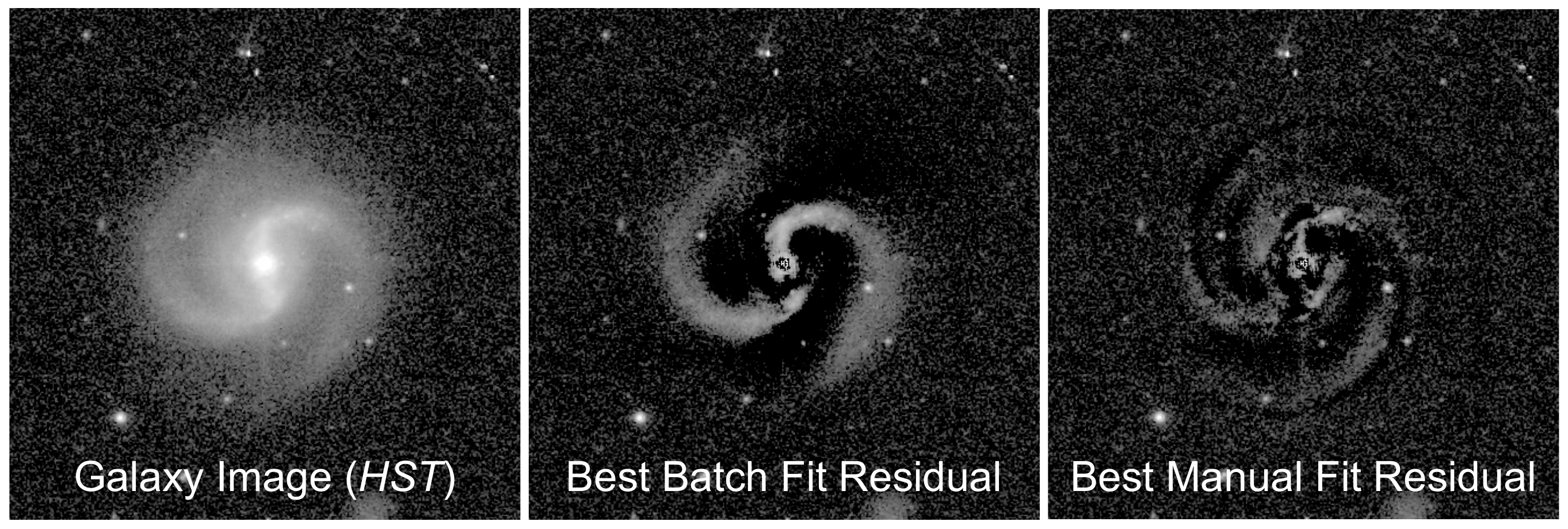}
    \caption{Galaxy J003432.51+391836.1 as it goes through the fitting procedure. The first panel shows the galaxy in the F814W filter prior to fitting. The second panel shows the best \galfit\ residual after undergoing the batch fitting process. The third panel shows the best \galfit\ residual after undergoing the final manual fitting procedure. All companion galaxies not included within the fit are masked so as not to contaminate the fit.}
    \label{fig:GALFITexample}
\end{figure}

Fig. \ref{fig:GALFITexample} shows galaxy J003432.51+391836.1 as it goes through the fitting procedure. The best model from the batch fitting procedure (residuals in Panel (b)) contains three components: a PSF to model the nuclear emission, a disk component with fixed S\'{e}rsic index ($n = 1$), and an additional S\'{e}rsic component with a free S\'{e}rsic index that appears to model a small central pseudobulge -- we discuss the distinction between classical bulges and pseudobulges in Section \ref{sec:ClassicalVPseudo}. During the manual fitting process, two additional S\'{e}rsic components are added for a total of five components including the PSF. One of these additional components models the bright spiral arms visible in the batch fit residual, and the other models a short central galactic bar also present in the batch fit residual. The residual from this manual fit is displayed in Fig. \ref{fig:GALFITexample}c.

During the manual fitting process, we allow component parameters to vary in as many cases as possible in order to ensure that the fit converges to a local minima (\ie\ is a `stable' fit). As every subsequent component is added, we fix the previous components to their best fit. Once all components are added, we re-run the fit one final time, allowing all components to vary, with their individual best fits as initial values. This ensures that the most robust and accurate component parameters are reported. The sky background is allowed to vary during this manual fitting process in order to ensure the extended regions of the galaxy are well constrained.

To obtain surface brightness (SB) values for each component at the effective radius (\reff), the fit is re-run a second time. All parameters other than SB for each component are fixed to those determined during the manual fitting procedure, and only SB is allowed to vary. For five galaxies, only SB values for the bulge components could be recovered, and thus the additional components in these fits only have full integrated magnitudes.

After fitting, eight galaxies are removed from the sample due to having morphologies consistent with elliptical galaxies, leaving us with 92 galaxies for further analysis.
Given that previous work has been reliant upon SDSS imagery and visual identification, it is encouraging that this work's in-depth fitting shows that these previous methods are robust enough to select small samples of disk-dominated galaxies at $\gtrsim 90$ per cent confidence.

In the final sample of 92 galaxies, every galaxy is fit with at least three components including the PSF, and most galaxies (71) are fit with at least four components.
A smaller number of galaxies (24) require five components, and only a single galaxy in our sample is fit with six components.

\subsection{Component Classification} \label{sec:ComponentClassification}
In order to quantify the luminosity contribution of each component to the host galaxy luminosity (\eg\ bulge contribution, pseudobulge contribution), each individual component within a given galaxy must be assigned a classification. This initial classification was carried out by a single expert classifier (MJF).

Each component is assigned a morphology: ‘disk’, ‘bulge’, ‘spiral arms’, ‘bar’, or ‘companion’, based on the final S\'{e}rsic index of a given component as well as its effective radius. Components with large radii and low S\'{e}rsic indices are assigned as ‘disk’, and components with small radii and high S\'{e}rsic indices are assigned as ‘bulge’.
In order to not underestimate the bulge contribution, ‘bulge’ is also assigned to more compact central galaxy components. At this stage, no attempt is made to distinguish between classical bulges, and visually similar pseudobulges. `Spiral arms' are identified through the presence of `spiral inner/outer radius' parameters in the final fit for that component, and `bar' is identified by the presence of `truncation' parameters in their angle and radius.

\subsubsection{Classical bulges versus pseudobulges} \label{sec:ClassicalVPseudo}

Pseudobulges are dense central components, making them almost visually indistinguishable from classical bulges. However, they are far more disk-like in their stellar dynamics, likely being created through secular growth processes by galaxy disks from galaxy disk material \citep{kormendy2004}. Since we aim to investigate the role of mergers in AGN evolution, it is important to accurately separate these secularly evolved pseudobulges from classical bulges.

We use the Kormendy relation \citep{kormendy1977, hamabe1987}, which describes the relationship between \reff\ and surface brightness measured at \reff\ for elliptical galaxies, to distinguish between classical bulges and pseudobulges. Surface brightness and \reff\ tend to be better constrained than S\'{e}rsic index. Components on or above the relation are likely dispersion-supported (\ie\ ellipticals, classical bulges). Components below the relation are likely rotation-supported (\ie\ disks, pseudobulges). Following \citet{nigoche-netro2008}, we choose the parameters of the Kormendy relation corresponding to the magnitude range of our sample in the SDSS $i$- and $z$-bands, $-20.0 > \mathcal{M} > -25.0$.

\begin{figure}
    \centering
    \includegraphics[width=\columnwidth]{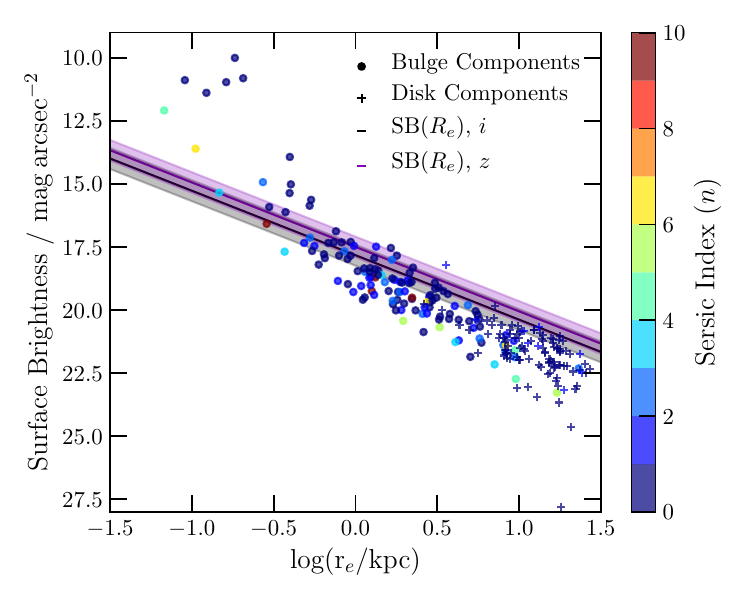}
    \caption{Surface brightness at the effective radius as a function of the effective radius for each of the components initially assigned a `disk' or `bulge' classification in our sample for the 92 galaxies identified as disk-dominated. `Disk' components are shown with crosses, while `bulge' components are assigned circles. The S\'{e}rsic index of a given component is indicated using the colour bar. Kormendy relations for the SDSS $i$- and $z$-bands are shown with the solid black and purple lines respectively. Shaded regions represent 1$\sigma$ confidence intervals.}
    \label{fig:KormendyRelation}
\end{figure}

The position of components initially classified as `disk' or `bulge' relative to the Kormendy relation is shown in Fig. \ref{fig:KormendyRelation}. Our chosen Kormendy relations for the SDSS $i$- and $z$-bands are shown with solid black and purple lines respectively, with the shaded region indicating the 1$\sigma$ confidence intervals.

Any `bulge' component lying below the lower 1$\sigma$ confidence interval of the SDSS $i$-band relation is designated a `pseudobulge', and any `bulge' component which lies on or above this line is considered a true classical bulge and is designated a `bulge'. We choose this conservative threshold in order to minimise the chances of underestimating the contributions of classical bulges to our sample. We find that 49 galaxies with our sample ($53.3 \pm 0.5$ per cent) host a classical bulge. This errs on the side of misclassifying a pseudobulge as a classical bulge, rather than vice-versa, in order that we can select a very pure sample of pseudobulge-only disk galaxies.

The majority of `bulge' and `disk' components lie below the Kormendy relation, implying that most of the components in our sample are rotation-supported structures. This confirms that our sample consists of overwhelmingly disk-dominated morphologies. In addition, we find that true bulge components within the sample are small compared to pseudobulge and disk components, with an average \reff\ of $1.33\kpc$ at the median redshift of our sample ($z = 0.129$).

Previous studies have often used exclusively the S\'{e}rsic index of a component to distinguish between classical bulge and pseudobulge, with a typical value of $n = 2$ found to be a reliable metric for characterising a large sample \citep{fisher2008}. However, Fig. \ref{fig:KormendyRelation} shows that the S\'{e}rsic index alone is not a reliable indicator of the true nature of a bulge-like component within our sample, consistent with studies such as \citet{gadotti2008}. \citet{graham2001} demonstrated that bulges in spiral galaxies can have a wide range of S\'{e}rsic indices. We find 35 components designated as classical bulges with measured S\'{e}rsic indices in the range $0 \leq n \leq 1 $, and 9 components designated as pseudobulges with S\'{e}rsic indices $n \gtrsim 4$. This further underlines the importance of parametric fitting techniques to accurately classify galaxy components.

\citet{gadotti2008} note that the structural parameters of bulges (in particular, bulge S\'{e}rsic index) are profoundly affected by improper modelling of luminous AGNs or galactic bars. Whilst every care has been taken during our fitting procedure to accurately model the AGN through the PSF and to fit accurate galactic bar components, our sample is somewhat extreme when compared to the general galaxy population, with small bulges in comparison to dominant disks, combined with very bright central AGN. Our use of the Kormendy relation for bulge classification aids in mitigating these issues. Whilst individual S\'{e}rsic indices may be poorly constrained, utilising both \reff and SB of the components allows us to more accurately determine the morphological nature of a given component.

Throughout the rest of this work, we use `bulge' to refer exclusively to classical bulges.

\section{Additional sample parameters} \label{sec:AdditionalParameters}

We obtain accurate estimates of luminosity ratios, galaxy stellar masses (\mstar), bulge stellar masses (\mbulge), black hole masses (\mbh), AGN luminosities (\lbol), and Eddington ratios (\fedd). By comparing the distributions of these parameters for a bulge-only sample, and a pseudobulge-only sample, we can determine how galaxy and black hole growth relate in the absence of major mergers.

\subsection{Luminosity Ratios} \label{sec:LRatios}

Having carried out the component classification detailed in Section \ref{sec:ComponentClassification} we compute AGN-to-total, bulge-to-total, and pseudobulge-to-total luminosity ratios for each galaxy in the sample.

The host galaxy luminosity is taken to be the sum of the luminosity of each component used to fit the galaxy, \emph{excluding} the PSF which represents the nuclear (AGN) emission. When computing AGN-to-total ratios we assume the PSF luminosity represents the entirety of the AGN luminosity, and that any contribution from the host galaxy is minimal. The bulge and pseudobulge luminosities are taken to be the sum of the relevant component luminosities.

The AGN-to-total luminosity ratio (\agntotL) for the sample lies between $0.003<\magntotL<1.6$, with a mean value of $0.2 \pm 0.2$, and a median value of $0.19 \pm 0.03$. This confirms the highly luminous nature of the AGNs in the sample, and the importance of robustly constraining the PSF when engaging in structural decomposition of AGN host galaxies.

For galaxies with detected bulges, the bulge-to-total luminosity ratio (\btotL) lies between $0.04<\mbtotL<0.52$, with a mean value of $0.2 \pm 0.1$ and a median value of $0.18 \pm 0.02$. When we include galaxies with no detected bulge component, the mean and median \btotL\ for the sample drops to $0.1 \pm 0.1$ and $0.06 \pm 0.03$ respectively.

\citet{simard2011} carried out bulge-disk structural decomposition of 1.12 million galaxies in the SDSS DR7 Legacy survey, determining best-fit models and structural parameters for each source.

\citetalias{simmons2017} use these bulge-disk decompositions of the 90 galaxies present in both their own and the \citet{simard2011} sample to obtain an upper limit \btotL\ in the $r$-band. These lie between 0.13 $\leq \mbtotL\ \leq$ 1.0, with a mean value of 0.5. In addition, \citetalias{simmons2017} constrain 2 of the sources imaged with the INT to have \btotL\ of $0.3 \pm 0.2$ and $0.47 \pm 0.2$.

\begin{figure}
    \centering
    \includegraphics[width=\columnwidth]{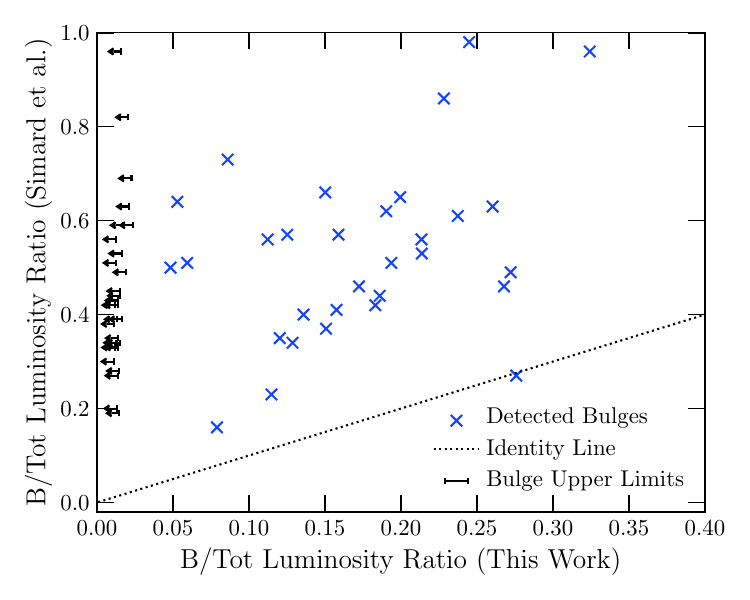}
    \caption{Recovered bulge-to-total luminosity ratios (\btotL) for the 59 galaxies from our disk-dominated sample that also appear in the \citet{simard2011} catalogue. Galaxies from our sample with detected bulge components are shown with blue crosses, whilst galaxies with no detected bulge components have the upper limits on their \btotL\ indicated with black arrows. To aid visual clarity, these limits are given an additional +0.01 shift along the $x$ axis. The black dotted line represents the Identity Line ($y = x$). $>98$ per cent of these galaxies have reported \btotL\ in \citet{simard2011} that are higher than those found by this work, demonstrating the extent to which bulge contribution to galaxy luminosity can be artificially inflated if a bright AGN is not accounted for during bulge-disk decomposition.}
    \label{fig:SimardComp}
\end{figure}

We similarly match our galaxy catalogue to \citet{simard2011}, and find that 59 galaxies appear in both catalogues. Following \citetalias{simmons2017}, we find \btotL\ for these 59 galaxies in the range 0.16 $\leq \mbtotL \leq$ 0.98, with a mean value of 0.5. Fig. \ref{fig:SimardComp} shows a comparison of the recovered \btotL\ of our galaxies versus those recovered by  \citet{simard2011}. 

Our mean \btotL\ for this subsample, including galaxies with no detected bulge component, lies below even the smallest upper \btotL\ limit of the \citet{simard2011} galaxies. When we consider only those galaxies with detected bulges, the mean \btotL\ for our sample lies below both the mean \btotL, and each of the individually constrained \btotL\ from the \citet{simard2011} sample, strongly indicating that the overall bulge contribution to this sample is small, despite $> 50$ per cent of the galaxies in the sample containing a classical bulge component (see Section \ref{sec:ClassicalVPseudo}).

For galaxies with detected pseudobulges, the pseudobulge-to-total luminosity ratio (\pbtotL) ranges from $0.03 <\mpbtotL< 0.96$, with a mean value of $0.3 \pm 0.3$, and a median value of $0.2 \pm 0.3$. When we include galaxies with no detected pseudobulge component the mean and median \pbtotL\ for the sample drop to $0.2 \pm 0.2$ and $0.09 \pm 0.03$ respectively. As such, we find in either case that the mean \pbtotL\ are comparable to, if not larger than, the the mean \btotL\ reported above, indicating the increased prevalence of pseudobulge components within the sample. When we consider the subset of galaxies ($23.9 \pm 0.2$ per cent) in our sample which contain both pseudobulges and classical bulges, we find mean \pbtotL\ and \btotL\ of $0.3 \pm 0.2$ and $0.2 \pm 0.1$ respectively. As such, even when considering compact central components in our galaxies, the majority of components are rotation dominated in nature.

Furthermore, Fig. \ref{fig:SimardComp} demonstrates the effect of not taking into account a bright AGN when performing galaxy bulge-disk decomposition, with the bulge contribution to galaxy luminosity significantly artificially inflated. Considering only the galaxies with detected bulges we find that the \citet{simard2011} \btotL\ are larger than those recovered in this work by factor of $\sim3.75$ on average, and we find some \btotL are larger by a factor of $\geq10$.

When considering the highly luminous nature of our AGNs in conjunction with the classical bulges being typically very small, it is possible that some true bulge components are not recovered due to the presence of the luminous AGN. In order to minimise the chances of underestimating the contributions of classical bulges, we carry out the following procedure on all galaxies regardless of whether or not a bulge component is detected.
 
The full width at half-maximum (FWHM) of the PSF of \HST-ACS is 2.5 pixels. As such, the largest bulge component that could be `hidden' underneath the PSF in our sample would have effective radius of 1.25 pixels, subtending an angle of 0.0625\arcsec\ on the sky. We compute the size of such a bulge in kiloparsecs for each of our galaxies, and then use the Kormendy Relation for the SDSS $z$-band (see Fig. \ref{fig:KormendyRelation}) to obtain an estimate of the surface brightness at the effective radius for each of these objects. We choose the $z$-band relation as this gives us the brightest possible value of surface brightness, and prevents the contribution from any hidden bulge component being underestimated. We use this surface brightness to calculate the luminosity of this extra hidden bulge component.
 
For galaxies with no detected bulge component, the luminosity ratio of this hidden bulge to the total luminosity is considered an upper limit of the \btotL\ for that object. These upper limits lie between $0.001 <\mbtotL<0.013$, with a mean value of $0.006 \pm 0.004$. For galaxies with a detected bulge component we utilise this extra luminosity in conjunction with the already calculated errors on the luminosity of the detected bulge components to inform the upper uncertainties on our \btotL\ for these galaxies.

This also shows that distance has very no practical impact on the fits. Fig. \ref{fig:SimardComp} shows that the maximum luminosity of the bulgeless components treating the limits as detections is significantly smaller than the subsample with detected bulges.

\subsection{Galaxy and Bulge Stellar Masses} \label{sec:StellarMasses}
To calculate stellar masses of the components we use the well-studied relation from \citet{baldry2006} between stellar mass, absolute galaxy $r$-band magnitude, and $u - r$ galaxy colour, \citep[corrected for galactic extinction,][]{schlegel1998}. To obtain photometry for our sample we perform a positional match to SDSS DR16 \citep{ahumada2020}. For one galaxy no photometric data is available in DR16, and so we use photometry from SDSS DR9 \citep{ahn2012}. In order to better remove the AGN contribution to galaxy luminosity, we correct the SDSS $u$- and $r$-band magnitudes for PSF contribution by multiplying the flux from the SDSS \texttt{modelMag} by the host-to-total luminosity ratio, \htotL\ computed using our fits from \galfit. We take the host galaxy luminosity as the sum of each galaxy component \emph{excluding} the PSF. The total galaxy luminosity is the sum of each galaxy component \emph{including} the PSF.

\begin{figure*}
    \centering
    \includegraphics[width=\textwidth]{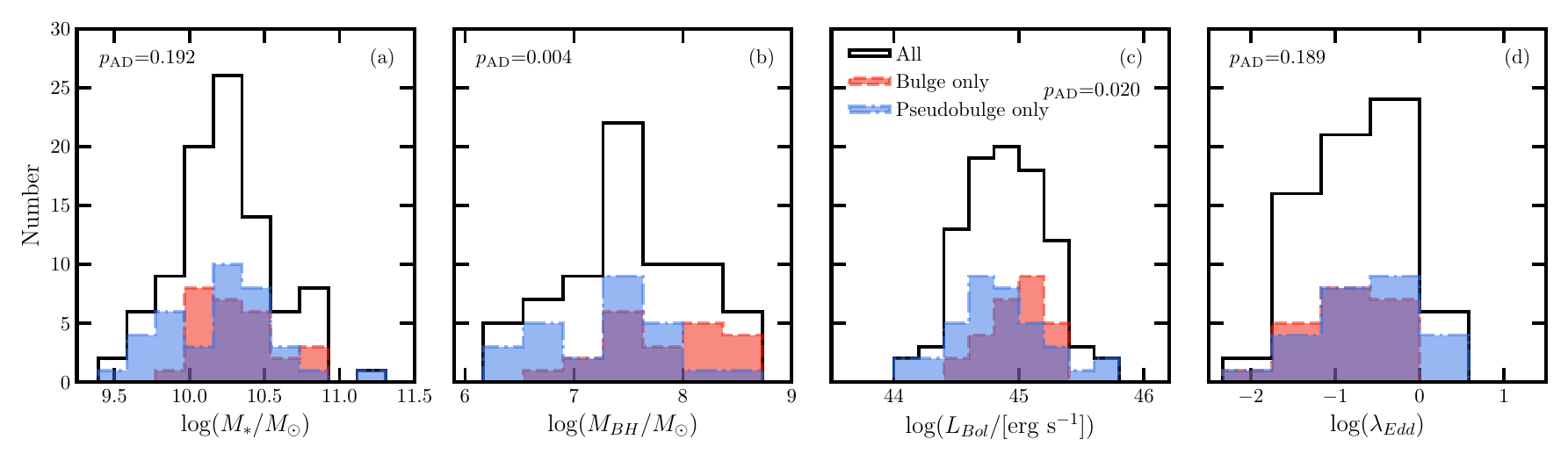}
    \caption{Distributions of galaxy stellar mass, \mstar\ Panel (a), black hole mass, \mbh\ Panel (b), bolometric luminosity, \lbol\ Panel (c), and Eddington fraction, \fedd\ Panel (d). Results for the entire sample are in black, for galaxies which contain only bulge components and no pseudobulge in dashed red, and for galaxies which contain only pseudobulge components and no bulge in blue hatched. Labels in the top left report $p$ values for Anderson-Darling tests when comparing the bulge only and pseudobulge only samples. For the \lbol\ and \mbh\ distributions we reject the null hypothesis that the bulge only and pseudobulge only samples are drawn from the same population, with $p_{\mathrm{AD}} = 0.020$ and $p_{\mathrm{AD}} = 0.004$ respectively. For the distributions of \mstar, and \fedd\ these two samples are more similar and the null hypothesis cannot be confidently ruled out. The resultant $p_{\mathrm{AD}}$ and $p_{\mathrm{KS}}$ are shown in Table \ref{tab:PVals}}
    \label{fig:SampleParameters}
\end{figure*}

We recover galaxy stellar masses (\mstar) ranging from $2.8 \times 10^{9}  < \mmstar/\mmsun < 8.4 \times 10^{10} $. The median \mstar\ is $1.7 \times 10^{10} \mmsun$. Each individual mass is taken to have an uncertainty of 0.3 dex from the scatter in the colour-luminosity relation. Fig. \ref{fig:SampleParameters}c displays the galaxy stellar mass distribution for our sample.

\begin{table}
    \caption{Tests for statistical significance performed on the bulge only and pseudobulge only distributions from Fig. \ref{fig:SampleParameters}. $p$ values are reported for both Kolmogorov-Smirnov (KS) and Anderson-Darling (AD) frameworks. For \mstar, and \fedd, the two samples are consistent with being drawn from the same parent sample. For \lbol\ and \mbh, the two samples are inconsistent with being drawn from the same parent sample, and we find AD statistics of 2.97 and 4.76, indicating statistically significant differences in these populations to 2.3$\sigma$ and 2.9$\sigma$ respectively.}
    \label{tab:PVals}
    \begin{tabular}{lcc}
    \hline
         & $p_{\mathrm{KS}}$ value & $p_{\mathrm{AD}}$ value \\ \hline
       Galaxy Stellar Mass (\mstar) & 0.099 & 0.192 \\
       Black Hole Mass (\mbh) & 0.012 & 0.004 \\
       AGN Bolometric Luminosity (\lbol)  &  0.016 & 0.020 \\
       Eddington Fraction (\fedd) & 0.384 & 0.189 \\
       \hline
    \end{tabular}
\end{table}

We report significantly lower \mstar than \citetalias{simmons2017} measurements of $8 \times 10^{9} <\mmstar/\mmsun<2 \times 10^{11} $, with a mean galaxy stellar mass of $4 \times 10^{10} \mmsun$. We attribute these differences primarily to changes in the SDSS pipeline from DR8 to DR16, which have particular effect on recovering galaxy magnitudes in the $u$-band.

Fig. \ref{fig:SampleParameters}a also displays the M* distributions for the bulge-only and pseudobulge-only galaxies separately.  The results of Kolomogrov-Smirnov (KS) and Anderson-Darling (AD) tests for the statistical significance of the difference in these distributions are listed in Table \ref{tab:PVals}. We find $p$ values of $p_{\mathrm{KS}} = 0.099$ and $p_{\mathrm{AD}} = 0.192$ when comparing the bulge-only (\ie\ no pseudobulge component) and pseudobulge-only (\ie\ no classical bulge component) \mstar, indicating the two samples are consistent with being drawn from the same parent sample.
Whilst at high and median masses the distribution of the bulge-only and pseudobulge-only samples are very comparable, the bulge-only sample seems to lack the same low-mass tail displayed by the pseudobulge-only sample, dropping off rapidly at $\sim10^{10} \,\mmsun$. This could indicate that the galaxies in our sample which host bulges tend to have stellar masses above a certain mass threshold, with few galaxies at low stellar mass.

These findings are consistent with \citet{skibba2012}, who show a strong correlation between bulge strength and galaxy stellar mass in their sample of $\sim 16,000$ disk galaxies at $z < 0.06$. Most strikingly, the transition from disk-dominated to bulge-dominated for this correlation appears at $\mmstar \approx 2 \times 10^{10} \mmsun$, very close to the threshold in our sample below which galaxies hosting classical bulges are mostly absent. Such findings could indicate that these bulge-only galaxies are present in different environments to those of the pseudobulge-only sample. Higher density environments might allow for better accretion of cold molecular gas and lead to increased star formation, as well as increasing the likelihood of galaxy mergers (major or minor) explaining why these objects which host bulges also seem to possess stellar masses above a given threshold. This is consistent with the findings of \citet{skibba2012} who report their bulge-dominated disk galaxies reside in denser environments.

Further work to investigate the star formation rates of these objects, as well as their gas supply and local environment might shed further light on the extent to which these are affecting long term star formation in our bulge-only sample. We highlight here that when we also consider the $p_{\mathrm{AD}}$ value we find that the bulge-only and pseudobulge-only samples are consistent with being drawn from the same parent sample.

We use our luminosity ratios (see Section \ref{sec:LRatios}) in conjunction with our galaxy stellar masses in order to obtain estimates of bulge stellar mass, \mbulge\ for each of our galaxies. These correspond to \mbulge\ (for those galaxies with detected bulges) in the range $3.9 \times 10^{8} <\mmbulge/\mmsun<1.8 \times 10^{10}$, with a mean \mbulge\ of $3 \times 10^{9} \mmsun$.

When we consider the upper limits on \btotL\ for those galaxies with no detected bulge contribution (see Section \ref{sec:LRatios}) we find they correspond to \mbulge\ in the range $1.7 \times 10^{7} < \mmbulge/\mmsun < 4.4 \times 10^{8}$, with a mean \mbulge\ of $7.4 \times 10^{7} \mmsun$.

\subsection{Black Hole Masses} \label{sec:BHMasses}

\citetalias{simmons2017} use the relationship from \citet{greene2005} \citep[subsequently re-calibrated by][]{shen2011} to obtain black hole masses, \mbh, from the established relationship between \mbh, and FWHM and luminosity in the broad H$\alpha$ line. This relation is chosen to avoid contamination of the spectra by the host galaxy, and because of H$\alpha$'s availability for all sources in this sample. A bootstrap method, sampling each value $10^{5}$ times within the uncertainties was used by \citetalias{simmons2017} in order to estimate the uncertainties on these black hole masses. As such, we recover black hole mass estimates for 68 of the galaxies in our sample. We refer to \citetalias{simmons2017} for further information regarding the calculation of black hole masses.

\mbh\ for galaxies in this subsample range from $1.8 \times 10^{6} <\mmbh/\mmsun<4.8 \times 10^{8}$. The median \mbh\ is $3.0 \times 10^{7} \mmsun$.

Fig. \ref{fig:SampleParameters}b shows the \mbh\ distribution for the 68 galaxies with available masses in our sample. The bulge-only galaxies within our sample lie at the higher mass end, with a median \mbh\ of $7.3 \times 10^{7} \mmsun$ compared to a median \mbh\ of $2.4 \times 10^{7} \mmsun$ for the pseudobulge-only population.

When comparing the bulge-only and pseudobulge only \mbh, the $p$ values of $p_{\mathrm{KS}} = 0.012$ and $p_{\mathrm{AD}} = 0.004$ suggest that the two samples are inconsistent with being drawn from the same parent sample, to $2.9\sigma$ in the case of the AD results. This suggests that, within our sample, galaxies which host bulges also host more overly massive SMBH in comparison to those galaxies which do not have a bulge component.

We note that \mbh\ are of the same order of magnitude regardless of bulge or pseudobulge, indicating that whilst merger events may result in increased \mbh, they are likely not the dominant mechanism facilitating this. Simulation work \citep{martin2018, McAlpine2020, smethurst2024} suggests that mergers play little role in SMBH growth in the long term. This point is further highlighted when we consider those galaxies in the sample with neither a pseudobulge nor a classical bulge, which present with a median black hole mass of $2.9 \times 10^{7} \mmsun$. Such black hole masses are highly comparable to the overall sample and easily breach $10^{7} \mmsun$, challenging previous notions that galaxies with no pseudobulge or bulge contribution should struggle to grow their black holes above $\sim 10^{5-6} \mmsun$ \citep{satyapal2008, satyapal2009, secrest2012}. These findings are in agreement with more recent literature that has also confirmed black holes grown to supermassive size through secular processes, independent of major mergers \citep{bohn2020, bohn2022}.

Whilst galaxies with a recent history of mergers experience enhanced black hole growth, our merger-free galaxies can also grow their black holes to significant and comparable masses in the local Universe. However, we note here that our sample preferentially selects the most luminous AGNs, which are likely to have more massive black holes. As such, these truly bulgeless galaxies in our sample with black holes $\geq10^{7} \mmsun$ may represent the extreme tail of a distribution of SMBHs fuelled by purely secular growth. Whilst it is possible to grow black holes to masses in excess of $\sim 10^{5-6} \mmsun$ through purely secular processes, such black holes may not be necessarily representative of the global secularly-grown black hole population. Further study of more moderate luminosity AGNs hosted in disk-dominated galaxies would aid in better understanding the overall distribution of black hole masses in these systems, and the frequency at which such black holes grow to $\geq10^{7} \mmsun$.

\subsection{AGN Luminosities, Accretion, and Eddington Ratios} \label{sec:Accretion}

We estimate bolometric AGN luminosities, \lbol, for our sample following \citetalias{simmons2017} utilising the \textit{WISE} W3 band centred at $12\micrometer$. We choose a bolometric correction factor of $\approx 8$ from \citet{richards2006}, which does not depend significantly on wavelength in the infrared (see their Fig. 12). This results in a \lbol\ range of $1.1 \times 10^{44}\leq \mlbol/ \erg\s^{-1} \leq5.5 \times 10^{45}$. The median luminosity is $7.7 \times 10^{44} \erg \s^{-1}$.

Fig. \ref{fig:SampleParameters}a shows the \lbol\ distribution for our sample. We find $p$ values of $p_{\mathrm{KS}} = 0.016$ and $p_{\mathrm{AD}} = 0.020$ when comparing the bulge-only and pseudobulge-only samples (Tab. \ref{tab:PVals}), with the bulge-only sample having slightly higher luminosities. These $p$ values show that the two samples are inconsistent with each other, to $2.3\sigma$ in the case of the AD results. This suggests that galaxies which host bulges also host more luminous AGNs.

Observational studies show that luminous quasars are preferentially hosted in ongoing mergers \citep{volonteri2006, urrutia2008, glikman2015, trakhtenbrot2017}, and simulations have revealed that mergers increase the rate of luminous AGNs \citep{McAlpine2020}. This supports our increased incidence of luminous AGNs in the galaxies hosting bulges. When we consider these results in conjunction with our findings in Sections \ref{sec:StellarMasses} and \ref{sec:BHMasses}, these AGNs may display increased luminosity due to their higher mass SMBH and potentially more dense environments. However, the above referenced works concern galaxies engaged in ongoing mergers, whilst our sample has been specifically selected due to a lack of an apparent major merger event since $z \sim 2$, a lookback time of $\sim 10^{10}\yr$. Given that typical AGN duty cycles are on the order of $10^{5 - 6} \yr$ \citep{keel2012, schawinski2015}, it seems unlikely that merger events in the distant past are influencing their AGN luminosities in the present.

Our results from Sections \ref{sec:StellarMasses} and \ref{sec:BHMasses} may indicate that the galaxies within our sample reside in denser environments. Such environmental density might produce the effect of simultaneously facilitating more galaxy mergers resulting in an increased proportion of bulge components, and allowing for higher inflow of cool molecular gas, leading to enhanced star formation and increased total galaxy stellar mass. These two processes could be working in tandem to funnel more gas to the galactic centre resulting in a more luminous AGN with a more massive black hole. However, studies indicate that the unpredictable gas inflow geometry of merger driven fuelling should have the effect of spinning down any given black hole, resulting in lower radiative efficiency and more gas mass lost to outflows \citep{smethurst2020, beckmann2024}. This could have the combined effect of both limiting the gas available to grow a black hole, as well as reduce its bolometric luminosity.

\lbol\ is related to mass accretion rate, $\dot{m}$ through a simple matter to energy conversion, where $\eta = 0.15$ \citep{elvis2002} is the efficiency of conversion

\begin{equation}
    \dot{m} = \frac{\mlbol}{\eta c^{2}}
    \label{eq:accretion}
\end{equation}

\citetalias{simmons2017} use black hole masses to calculate the luminosities expected if each AGN were accreting at its Eddington limit, \ledd. We use this to measure the fractional growth rate of the black hole relative to the maximum rate it is capable of sustaining, via the Eddington fraction, \fedd.

\begin{equation} \label{eq:REdd}
    \mfedd \equiv \frac{\mlbol}{\mledd}
\end{equation}

Fig. \ref{fig:SampleParameters}d shows the \fedd\ distribution for our sample. We find $p$ values of $p_{\mathrm{KS}} = 0.384$ and $p_{\mathrm{AD}} = 0.189$ when comparing bulge-only and pseudobulge-only samples (Tab. \ref{tab:PVals}), indicating that the two samples are consistent with being drawn from the same parent sample. This suggests that galaxies which host bulges do not have a statistically significantly higher fractional SMBH growth rate in comparison to those which have no bulge component.

This further supports the notion that whilst galaxy mergers might enhance black hole growth, merger events are not the only process facilitating this growth. Such highly comparable Eddington fractions imply that the dominant pathways fuelling black hole growth for both the bulge-only and pseudobulge-only subsamples are also comparable. This evidences a more fundamental process common to every galaxy in our sample as the dominant pathway for facilitating black hole growth in the long term. As mentioned previously, this view is consistent with simulation work from  \citet{martin2018} and \citet{McAlpine2020} who suggest that mergers contribute as little as $\sim 15$ per cent of black hole growth since $z \sim 3$.

\section{Bar fraction}\label{sec:BarFraction}
Galactic bars have been proposed as a secular mechanism for black hole feeding by transporting gas to the central AGN \citep{shlosman1989, friedli1993, athanassoula2003, ann2005}, although there is much disagreement in the literature. Many studies find no correlation between AGNs and bars \citep{martini2003, lee2012, cheung2015, goulding2017}, whilst others find that bars may be preferentially hosted in active galaxies with an overall incidence increase of $\sim 20$ per cent \citep{knapen2000, laine2002, laurikainen2004, oh2012, garland2023, garland2024, kataria2024}. \citet{galloway2015} note that whilst there is a higher probability of a galaxy hosting an AGN also hosting a bar, they find no link between bars and the quantity or efficiency of AGN fuelling, indicating that whilst bars may trigger AGNs they have little further effect once the AGN is established. \citet{garland2024} distinguish between weak and strong bars, and in doing so, find to a $>5\sigma$ confidence that strongly barred galaxies are more likely to host an AGN than weakly barred galaxies, which are more likely to host an AGN than unbarred galaxies. Investigating the bar fraction in our sample in detail with high-resolution \HST\ imaging could provide crucial information about the role of bars in fuelling AGNs.

Within our sample, 23 galaxies require a component modelling a galactic bar, representing $25 \pm 2$ per cent of our galaxies \citep[where errors arise from the binomial distribution,][]{cameron2011}. This is consistent with previous literature including \citet{masters2011} who found that around $29.4 \pm 0.5$ per cent of disk galaxies host a large scale galactic bar at redshift $0.01 < z < 0.06$ when imaged in the optical. Of the 23 galaxies hosting bars, 6 are completely bulgeless (no pseudobulge or classical bulge contribution), 10 host only a pseudobulge, 5 host only a classical bulge, and 2 host both a classical bulge and a pseudobulge.

The AGN bolometric luminosities for the barred galaxies range from $3.6 \times 10^{44} \leq \mlbol/\erg \s^{-1} \leq4.6 \times 10^{45}$, with a median bolometric luminosity of $7.8 \times 10^{44} \erg \s^{-1}$, in good agreement with the luminosity range of our overall sample (Section \ref{sec:Accretion}), and upon performing tests for statistical significance we find $p$ values of $p_{\mathrm{KS}} = 0.801$ and $p_{\mathrm{AD}} > 0.25$ indicating that the barred and non-barred galaxies have \lbol\ consistent with being drawn from the same parent sample.

\mbh\ estimates are available for 18 of the barred galaxies within our sample, ranging from $2.0 \times 10^{6} \leq\mmbh/\mmsun\leq1.8 \times 10^{8}$, with a median mass of $2.9 \times 10^{7} \mmsun$. Similarly to the AGN bolometric luminosity, the black hole masses are in very good agreement with the overall sample -- we find $p$ values of $p_{\mathrm{KS}} = 0.135$ and $p_{\mathrm{AD}} > 0.25$, also indicating that the barred and non-barred galaxies have \mbh\ consistent with being drawn from the same parent sample. This suggests that barred galaxies are hosting neither AGNs that are overly luminous, nor black holes which are overly massive when compared to our overall sample of disk galaxies.

If bars are the primary fuelling mechanism for black hole growth in the merger-free regime, we might expect that AGNs hosted in barred galaxies would be more luminous and have larger black hole masses when compared to unbarred galaxies. This is especially relevant when we consider that galactic bars are thought to be long-lived morphological features \citep{debattista2006, kraljic2012, geron2023}, and so any additional black hole growth they facilitate in the long term should be reflected in measured black hole masses in the local Universe. The fact that our findings show no tendency for overly massive black holes implies bars are simply one element of the overall picture, with some other mechanism common to all galaxies in our sample providing the majority of the fuel for black hole growth. Such findings are consistent with \citet{galloway2015}, who present a scenario wherein bar-driven fuelling contributes some fraction of the fuel for growing black holes, whilst some other processes must also contribute to black hole accretion disk fuelling through angular momentum transfer at smaller radii. This scenario is also consistent with simulations that show multiple large-scale mechanisms including galactic bars can be responsible for transporting gas to the scales required for AGN fuelling \citep{hopkins2010}. This suggests that the presence of a bar is not a requirement to grow black holes to supermassive size in the local Universe in the absence of major mergers.

\section{Galaxy-black hole mass relations} \label{sec:GalBHMassRels}

\begin{figure*}
    \centering
    \includegraphics[width=\textwidth]{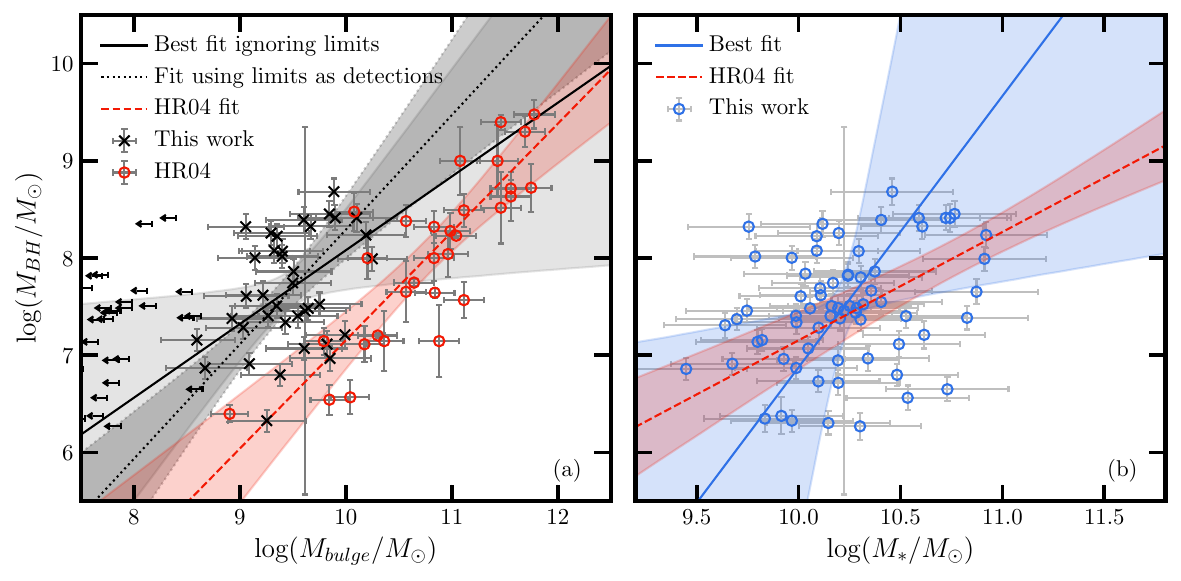}
    \caption{Panel (a), black hole mass versus bulge stellar mass. Black crosses indicate bulge masses for galaxies with detected bulge components. Arrows indicate upper limits on bulge masses for galaxies with no detected bulge component. The best fit to galaxies with detected bulges only is shown with a solid black line. The best fit incorporating the upper limits into a censored fit is shown with a dotted line. Red open circles show early-type galaxies and the best fit is indicated by the red dashed line. Shaded regions indicate 2$\sigma$ confidence intervals. Panel (b), black hole mass versus total stellar mass. Open blue circles show the stellar masses for our sample, the best fit to these is shown in the solid blue line. The \citetalias{haring2004} relation for early type galaxies  is again shown with the red dashed line, unchanged from the left panel.}
    \label{fig:GalBHRels}
\end{figure*}

We investigate the relationship between \mbh\ and \mbulge. Again, the BH-bulge relations of \citetalias{simmons2017} are constrained only to a very limited extent, as $\sim 98$ per cent of their bulge masses are upper limits. Our structural decomposition allows us to assign more accurate bulge mass estimates (Section \ref{sec:StellarMasses}) to a significant proportion of our sample.

We use a Bayesian method to fit a linear regression model to the 68 sources in our sample for which we have \mbh\ estimates from \citetalias{simmons2017}, including two-dimensional uncertainties. The results are displayed in Fig. \ref{fig:GalBHRels}a. We perform two fits to our data: one where we fit only to the \mbh\ of galaxies which have detected bulge components, and one where we incorporate the upper limits on \mbulge\ for those galaxies which have no detected bulge component into a censored fit. Considering the fit to galaxies only with detected bulge components, we find a Spearman correlation coefficient, $\mrhoS=0.35$, and a Pearson correlation coefficient, $\mrhoP= 0.37$ indicating some correlation, albeit with significant scatter. When we include the upper limits on \mbulge\ for those galaxies with no detected bulge component we find $\mrhoS = 0.41$, and a $\mrhoP = 0.44$. Hence, in both cases we find evidence for some correlation between \mbh\ and \mbulge, and this correlation is strengthened when including the upper limits on bulge mass. The slopes and normalisations for the fits are shown in Tab. \ref{tab:BHGalrelations}.

\begin{table}
    \caption{Equations for line of best fit shown in Fig. \ref{fig:GalBHRels}. Quoted errors are the $1\sigma$ errors.}
    \label{tab:BHGalrelations}
    \begin{tabular}{lcc}
    \hline
         & Slope & Normalisation \\ \hline
        \mbh-\mbulge, bulge only     & $0.762\pm 0.350$ & $0.464\pm 3.313$ \\
        \mbh-\mbulge, with bulgeless & $1.194\pm0.207$ & $-3.645\pm1.938$ \\
        \mbh-\mstar                  & $3.471\pm3.880$ & $-27.989\pm39.570$ \\
        \citetalias{haring2004}      & $1.114\pm0.157$ & $-3.984\pm1.624$ \\
       \hline
    \end{tabular}
\end{table}

For comparison, we adopt a similar procedure for a sample of early-type galaxies from \citet{haring2004} \citepalias[hereafter][]{haring2004}. These galaxies have morphologies indicating a history which includes major mergers. The fitted relation for the \citetalias{haring2004} sample is somewhat consistent with both of our fitted relations.
The \mbh-\mstar slopes from this work are in agreement to within 1 standard deviation (SD) with the \citetalias{haring2004} relation. When we treat the limits as detections, our normalisation in also in agreement with \citetalias{haring2004} to 1 SD, and when we consider only the galaxies with detected bulges, the normalisation is in agreement to 2 SD.
However, the \citetalias{haring2004} relation is significantly more strongly correlated, with $\mrhoS = 0.82$ and $\mrhoP=0.83$. 
As such, we can conclude that while there is some correlation between black holes and bulges in our disk-dominated sample, there is much stronger correlation between these two galaxy properties in bulge-dominated galaxies with rich merger histories, but overall, the relationships are consistent with each other.

The fact that the correlation is so much stronger for bulge-dominated galaxies and present but highly scattered for disk-dominated galaxies is interesting. It implies that processes which can grow both bulges and black holes (such as major galaxy mergers) are the dominant co-evolutionary pathways in these early-type galaxies, and regulate such growth much more rigidly. This explains why the early-type galaxies considered here have bulge masses that appear to grow in almost lockstep with their black holes. Disk-dominated galaxies on the other hand are engaged in these processes significantly less often, as evidenced by their lack of strong bulge components, and this may result in a more stochastic co-evolutionary history which is reflected in their highly varied bulge masses for a given black hole mass.

Previously understood pathways for merger driven co-evolution require that any observed black hole-galaxy co-evolution must be driven by merger events large enough to create appreciable bulge components. As such, the fact that there is weaker correlation between \mbulge\ and \mbh\ in our sample, which has been specifically selected to lack bulges, might be seen as unsurprising. However, if these merger driven pathways are indeed the primary mechanisms for black hole-galaxy co-evolution, then we would also expect the black holes of our bulgeless galaxies to be undermassive, especially in comparison to those from \citetalias{haring2004} with rich merger histories, which is not the case

In Section \ref{sec:BHMasses} we find that the black hole masses of the truly bulgeless galaxies (those with no bulge or pseudobulge component) lie extremely close to our parent sample distribution of disk-dominated galaxies, showing that black holes in truly merger-free systems can still grow to considerable mass. This point is further emphasised when we consider that the most massive black holes in our disk-dominated sample are of comparable mass to all but the most massive black holes (of $\sim10^{9}\mmsun$) in the bulge-dominated comparison sample of \citetalias{haring2004}.

Furthermore, Fig. \ref{fig:GalBHRels}a shows that, when we consider the galaxies with detected bulges, $\gtrsim 97$ per cent of them lie above (or to the left of) the correlation for bulge-dominated galaxies, and $\gtrsim 90$ per cent lie above (or to the left of) the 2$\sigma$ uncertainty region for that correlation. Even when we consider the upper limits on \mbulge\ for those galaxies with no fitted bulge component, we find that all of them lie above both the fitted relation of \citetalias{haring2004}, and the 2$\sigma$ uncertainty region, with a \mbh\ distribution that is consistent in breadth with both the galaxies in our sample with detected bulges, and the majority of the galaxies in \citetalias{haring2004}. These findings are indicative of considerably higher \mbh\ in our sample than those predicted by BH-bulge relations, even when we consider conservative upper limits on the \mbulge, consistent with \citetalias{simmons2017}.

The analysis we have carried out in Section \ref{sec:GalBHMassRels} thus far explicitly assumes that any detected bulge component in our disk-dominated sample has been formed and grown exclusively by merger-driven processes. However, recent observations have suggested that classical bulges may grow through processes other than major galaxy mergers.

\citet{guo2020} find that short, sub-kiloparsec radius bars form morphological components similar to classical bulges when they are destroyed, allowing for a secular channel of bulge formation. Furthermore, studies suggests that disk galaxy bulges could form from so-called `red nuggets' \citep{damjanov2009}. These objects present as ultra-compact spheroids at high redshift, with half-light radii on the order of $\sim 1 \kpc$. \citet{delarosa2016} confirm that compact core components of galaxies at $z\sim 0.1$ are structurally similar to red nuggets observed at $z\sim 1.5$, and that these components are not exclusive to massive elliptical galaxies. This implies that these objects could become the cores of modern day disk galaxies \citep{costantin2020}. This model is supported by simulation work showing that extended star-forming disks develop around red nuggets after compaction \citep{zolotov2015, tacchella2016}. Very compact bulge components at the cores of disk galaxies could form purely secularly, without the need for major galaxy mergers. When we consider our findings in Section \ref{sec:ClassicalVPseudo} that the average classical bulge component in our sample is small in radius, it is possible that these bulge components (or the smallest among them) were once red nuggets at high redshift, with the disk-dominated morphology we view today subsequently forming around them. Follow-up observations would allow us to investigate and constrain the dynamical properties of these small bulges.

\citet{bell2017} investigate galaxy stellar halos in order to explore galactic merger histories and find that bulges in nearby Milky Way-like galaxies have higher than expected masses if mergers were the dominant growth mechanism. Furthermore, \citet{park2019} and \citet{wang2019} both emphasise the importance of disk star migration in disk-dominated galaxies. Stars with orbits aligned with the galactic disk continuously migrate towards the galactic centre, without the need of merger-driven perturbations. \citet{park2019} suggest that as much as half of the spheroidal component of disk-dominated galaxies may arise from orbits aligned with the disk in this way. Given that the galaxies in our sample are disk-dominated with \mstar\ comparable to those of local Milky Way-like galaxies, disk--star migration pathways for bulge growth could be applicable to our sample, indicating that the \mbulge\ that we can attribute purely to merger-driven growth could be substantially less than the \mbulge\ shown in Fig. \ref{fig:GalBHRels}a.

Were this the case, then not only are the SMBHs in our sample overly massive compared to the canonical BH-bulge relation, but the bulges themselves are overly massive compared to canonical frameworks for predicting bulge sizes based upon mergers.

Fig. \ref{fig:GalBHRels}b shows the relationship between \mbh\ and \mstar. The fitted \citetalias{haring2004} relation is unchanged in Panel (b) relative to Panel (a). \citetalias{haring2004} do not perform bulge-disk decompositions for 29 of the 30 galaxies in their sample, and furthermore $\gtrsim 80$ per cent of these galaxies are given a visual classification in \citet{devaucouleurs1991} of type E or S0. Hence, we assume that $M_{\mathrm{Bulge}} \approx \mmstar$ for the galaxies in the \citetalias{haring2004} sample.

The slope and normalisation of our \mbh-\mstar relation are shown in Tab. \ref{tab:BHGalrelations}. The fitted lines for both our disk-dominated and the \citetalias{haring2004} bulge-dominated galaxies considered here are consistent with each other, however the uncertainties on our relationship are very large. Performing tests for correlation, we find $\mrhoS=0.31$, and $\mrhoP=0.32$. This indicates the presence of a correlation of similar strength to that found between \mbulge\ and \mbh\ for this sample. Such a correlation is weak, and this is evident from the large amount of scatter present in our stellar mass distribution.

Observing a correlation between \mstar\ and \mbh\ for both bulge-dominated early type galaxies and the disk-dominated galaxies of our sample, but with significantly more scatter for the disk-dominated galaxies has interesting implications for black hole galaxy co-evolution. \citet{kormendy2013} propose a scenario wherein black holes with $\mmbh \leq 10^{8.5}\, \mmsun$ at $z \leq 1.5$ exist primarily in disk-dominated galaxies where enough gas reaches the black hole to allow for modest AGN activity, but that this should be so stochastic in nature as to preclude the co-evolution of the black hole with its host galaxy. The galaxies we consider in this sample host black holes with $\mmbh \leq 5 \times 10^{8} \,\mmsun$, and have been merger-free since $z \sim 2$, and so fit neatly into this picture. However, it is clear that the AGN activity in these objects is far from modest, and some secular process must be driving SMBH growth in these systems at appreciable rates. The significantly increased scatter in the fitted relations for our disk-dominated sample lends more credibility to the picture presented by \citet{kormendy2013} that these secular growth processes are highly stochastic in nature, but the presence of correlation between \mbh\ and \mstar\ for our sample, however weak, implies that these processes are not stochastic enough to prohibit co-evolution.

\cite{greene2020} investigate the relationships between \mbh\ and \mstar\ for large populations of early- and late-type galaxies. They find a clear offset in the normalisation of the fitted relations for the different galaxy subsamples, with late-type galaxies typically displaying lower stellar masses for a given black hole mass, and with more scatter. This scatter becomes more apparent at lower $\mmstar \lesssim 10^{11}\, \mmsun$, particularly for late-type galaxies. This could explain the observed scatter in \mstar\ for our sample of disk-dominated galaxies, and supports the picture we present above wherein secular growth processes in disk-dominated galaxies increase the stochasticity of black hole-galaxy co-evolution, but do not preclude it.

Another explanation for the large degree of scatter in our \mstar\ is the method used to calculate them (Section \ref{sec:StellarMasses}). In removing the PSF contribution to the flux in the $u$- and $r$-bands, we implicitly assume that the PSF contribution to galaxy flux is identical in both these bands. As the filters differ in central wavelength by $\sim 2700$\AA\ this assumption will clearly have significant impacts on our ability to recover accurate PSF-corrected filter magnitudes. However, the spectral energy distributions of Type 1 quasars should not vary significantly between these bands \citep[][Fig. 10]{richards2006}. Additionally, when we investigate the \texttt{PSFmag} values for the $u$ and $r$ filters we find broadly similar values, with mean apparent mags in SDSS DR16 of 18.47 and 17.66 respectively. These issues are minor compared to the inherent 0.3 dex scatter in the colour-luminosity relation.

Additional \HST\ imaging of our sample in a different filter would allow for a calculation of galaxy colour, analogous to the $u - r$ colour used to calculate mass-to-light ratios. \citet{taylor2011} make use of SDSS $g - i$ colours to estimate galaxy mass-to-light ratios to a precision of $\lesssim 0.1$ dex, a marked improvement over the 0.3 dex scatter from the relation we implement. Given that we already have imaging of these galaxies in filters roughly corresponding to the SDSS $i$-band, follow-up in a filter such as F475W would provide information corresponding to the $g$-band, allowing for more accurate estimation of galaxy mass-to-light ratios. Along with the well studied \HST-ACS PSF, the removal of the AGN component from these galaxies would be significantly more accurate. Works such as \citet{schombert2022} have also endeavoured to provide estimates of galaxy mass-to-light ratios that are additionally corrected for the presence of bulge and disk components, and such frameworks are invaluable when performing stellar mass estimates in samples such as ours.

Obtaining improved \mstar\ would allow further investigation into whether the scatter present in Fig. \ref{fig:GalBHRels}b is inherent and indicative of the galaxy co-evolution picture outlined above, or simply a result of our estimations. This latter option leaves open the possibility of stronger correlations between stellar mass and black hole mass, such as those found in \citetalias{simmons2017}.

\section{Summary}\label{sec:Summary}
By utilising high-resolution \HST\ imagery, we have performed comprehensive structural decomposition and analysis on an unambiguously disk-dominated (\ie\ merger-free) galaxy sample hosting AGNs. We summarise our key findings here.

The majority of the components making up these galaxies lie below the Kormendy relation, indicating that they are rotation-dominated in nature. The average bulge-to-total luminosity ratio is $0.1 \pm 0.1$ for the entire sample, indicating the extremely low prevalence of classical bulges and dispersion supported structures within our sample. We find a mean bulge mass for those galaxies with detected bulges of $3 \times 10^{9}\, \mmsun$.

Our findings indicate that whilst galaxy mergers might result in enhanced AGN luminosity, black hole mass and galaxy stellar mass, they are unlikely to be the dominant process impacting long-term black hole growth. Regardless of whether or not galaxies have indicators of being involved in a past merger, we show that their black holes can grow to significant mass in the local Universe. Further observational work, including an HI survey with an instrument such as the IRAM-30m telescope would help to better probe the nature of the environments around these galaxies and determine the effect this may have on the sample parameters mentioned above.

Our sample consists of the most luminous AGNs, and as such these objects may represent the extreme tail of a distribution of black holes fuelled by primarily secular growth processes. As such, follow up work should aim to probe the more moderate luminosity regime, giving us a better overall picture of the extent to which purely secular growth processes fuel black hole growth in the long term.

We advocate for the construction of larger samples of local, bulgeless, disk-dominated galaxies for structural decomposition with high-resolution imagery. This would allow for further investigation of disk-dominated galaxies that still host small classical bulges at their centres, and determine whether our findings hold true for the global AGN population. Furthermore, such investigations may provide more detail on the dynamical nature of these small bulges, and reveal whether a significant fraction of them were once `red nuggets', or whether they have indeed formed through galaxy mergers at high redshift. This could shed light on the extent to which secular processes have grown classical bulges.

With its high-resolution, the \Euclid\ mission is in a unique position to identify the sources described above and improve our catalogue size for structural decomposition.

\section*{Acknowledgements}

ILG has received the support from the Czech Science Foundation Junior Star grant no. GM24-10599M.
BDS acknowledges support through a UK Research and Innovation Future Leaders Fellowship [grant number MR/T044136/1].
EG acknowledges the generous support of the Cottrell Scholar Award through the Research Corporation for Science Advancement.
EG is grateful to the Mittelman Family Foundation for their generous support.
RJS gratefully acknowledges support through the Royal Astronomical Society Research Fellowship.
MRT acknowledges support from an STFC Ph.D. studentship [grant number ST/V506795/1].
The Dunlap Institute is funded through an endowment established by the David Dunlap family and the University of Toronto.

The authors would like to thank C. Peng and the \galfit\ News and Discussion Forum for their guidance and support.

In addition to the computer software already cited in the above manuscript, the authors have benefited immensely from the publicly available programming language \texttt{Python}, including \texttt{NumPy} \citep{vanderwalt2011}, \texttt{SciPy} \citep{jones2001}, \texttt{Matplotlib} \citep{hunter2007} and \texttt{astroPy} \citep{robitaille2013}. We also made extensive use of NASA's Astrophysics Data System (ADS), Cornell University's arXiv, the JavaScript Cosmology Calculator \citep{wright2006}, and the Tool for Operations on Catalogues and Tables \citep[\texttt{Topcat},][]{taylor2013}.

This research is in part based on observations made with the NASA/ESA Hubble Space Telescope, obtained at the Space Telescope Science Institute, which is operated by the Association of Universities for Research in Astronomy, Inc., under NASA contract NAS5-26555. These observations are associated with program HST-GO-14606. Support for program HST-GO-14606 was provided by NASA through a grant from the Space Telescope Science Institute, which is operated by the Association of Universities for Research in Astronomy, Inc., under NASA contract NAS5-26555.

Funding for the Sloan Digital Sky Survey IV has been provided by the Alfred P. Sloan Foundation, the U.S. Department of Energy Office of Science, and the Participating Institutions. SDSS acknowledges support and resources from the Center for High-Performance Computing at the University of Utah. The SDSS web site is \url{www.sdss.org}.

This publication makes use of data products from the Wide-field Infrared Survey Explorer, which is a joint project of the University of California, Los Angeles, and the Jet Propulsion Laboratory/California Institute of Technology, funded by the National Aeronautics and Space Administration.

\section*{Data Availability}
The data used in this work are available via Zenodo, with DOI 10.5281/zenodo.13959600. 


\bibliographystyle{mnras}
\bibliography{bibliography} 




\bsp	
\label{lastpage}
\end{document}